\documentclass[prl,twocolumn,showpacs,preprintnumbers]{revtex4}
\usepackage{rotate}
\usepackage{epsf}

\newcommand{\insertfig}[2]{\mbox{\epsfxsize=#1cm \epsfbox{#2.eps}}}
\newcommand{\Bx}{x_{\rm B}}

\begin{document}

\preprint{DOE/ER/40762-270 \ UMD-PP\#03-024}

\title{Exclusive electroproduction of lepton pairs as a probe of nucleon structure}

\author{A.V. Belitsky$^{1}$, D. M\"uller$^{2,1}$}
\affiliation{$^1$Department of Physics, University of Maryland,
             MD 20742-4111, College Park, USA \\
         $^2$Fachbereich Physik, Universit\"at Wuppertal,
             D-42097 Wuppertal, Germany}

\begin{abstract}
We suggest the measurement of exclusive electroproduction of lepton pairs as a
tool to study inter-parton correlations in the nucleon via generalized parton
distributions in the kinematical region where this process is light-cone dominated.
We demonstrate how the single beam-spin asymmetry allows to perform such kind of
analysis and give a number of predictions for several experimental setups. We
comment on other observables which allow for a clean separation of different
species of generalized parton distributions.
\end{abstract}

\pacs{11.10.Hi, 12.38.Bx, 13.60.Fz}

\maketitle

Since the discovery of a non-zero spatial extent of the proton in pioneering
measurements of electromagnetic form factors \cite{HofMcA51}, the exploration
of the hadron's internal structure in terms of quark-gluon degrees of freedom
was the subject of intensive theoretical and experimental studies. Until
recently, however, only information on one-dimensional slices of the nucleon
was extractable from hadronic observables: The spatial charge and magnetization
distributions from the form factors alluded to the above; the $x$-momentum
fraction space parton densities from inclusive reactions, e.g., deeply
inelastic scattering of leptons off hadrons \cite{Breetal69}. With the
realization of the power of exclusive reactions \cite{MueRobGeyDitHor94},
the opportunity of nucleon holography \cite{BelMul02} was brought to life;
namely, the simultaneous measurement of the longitudinal momentum fraction of
partons and their transverse localization within the area of resolution set
by the photon virtuality \cite{Bur00} (see also \cite{RalPir01,Die02}). The
corresponding hadronic characteristics, known as generalized parton
distributions (GPDs), intertwine the aforementioned conventional hadronic
functions. However, they also depend on an extra variable: the $t$-channel
longitudinal momentum fraction $\eta$, the so-called skewedness, which
changes the apportion of longitudinal momentum between the absorbed,
$(x - \eta)$, and created, $(x + \eta)$, partons. Thus, GPDs, contrary
to conventional Feynman's parton densities, give access to inter-parton
correlations rather than mere probabilities through the study of skewedness
dependence.

The pioneering experimental study of these new functions at HERMES \cite{Aip01},
CLAS \cite{Ste01} and HERA \cite{Adl01} has been done through the exclusive
leptoproduction of the real photon off the proton $\ell p \to \ell' \gamma p'$
at high momentum transfer. This process is sensitive to deeply virtual Compton
scattering (DVCS) amplitude (see Eq.\ (\ref{ComptonTensor})) parametrized in
terms of GPDs. The latter enter also in the leptoproduction of mesons, i.e.,
$\ell p \to \ell' M p'$. Single lepton and nucleon spin asymmetries
\cite{GouDiePirRal97,BelMulKir01} in the former case allow the direct measurement
of GPDs, however, in a very specific kinematics, namely, $x = - \eta$. This
restriction is a direct consequence of the reality of the outgoing photon. Such
a measurement is unable to constrain the spin sum rule involving the parton's
orbital momentum \cite{Ji96}. Here a processes is needed where the momentum
fraction varies independently of the skewedness. This can be achieved by the
relaxation of the reality condition for the photon in the final state, i.e., by
the measurement of lepton pairs produced from a timelike $\gamma$-quantum in
the elastic electron scattering $e p \to e' p' \ell \bar\ell$
\cite{BelMulKir01,GuiVan02}; see also Ref.\ \cite{BerDiePir01} for a related
reaction.

\begin{figure}[htb]
\begin{center}
\mbox{
\begin{picture}(0,160)(90,0)
\put(-5,0){\insertfig{9.4}{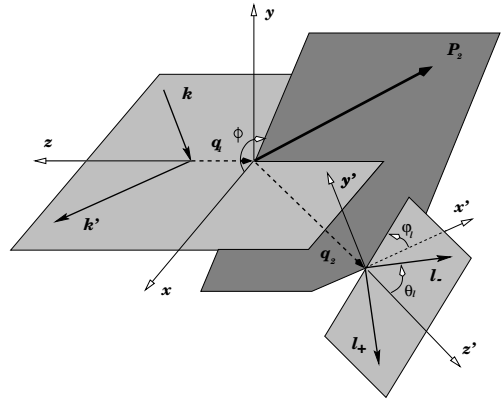}}
\end{picture}
}
\end{center}
\vspace*{-0.7cm}
\caption{\label{LeptonPairKinetic}
The kinematics of the lepton pair production in elastic electron scattering off
the proton, $e (k) p (p_1) \to e (k') p (p_2) \ell (\ell_-) \bar \ell (\ell_+)$,
in the fixed target setup.}
\end{figure}

\begin{figure}[tb]
\begin{center}
\mbox{
\begin{picture}(0,62)(115,0)
\put(-3,0){\insertfig{8.3}{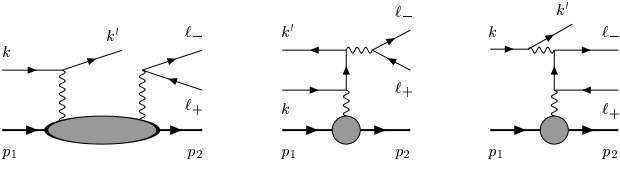}}
\end{picture}
}
\end{center}
\vspace*{-0.3cm}
\caption{\label{BHandDDVCS} Contributions to lepton pair production:
(from left to right) virtual Compton and Bethe-Heitler processes.}
\end{figure}

The differential cross section for the process in the kinematics displayed in
Fig.\ \ref{LeptonPairKinetic} reads
\begin{equation}
\label{Xsection}
d \sigma
= \frac{\alpha_{\rm em}^4}{16 (2 \pi)^3}
\frac{\Bx y}{{\cal Q}^2} |{\cal A}|^2
d \Bx d y d |\Delta^2| d \phi d M_{\ell \bar\ell}^2 d {\mit \Omega}_\ell
\, .
\end{equation}
The incoming photon virtuality and the invariant mass of the lepton pair are
$q_1^2 \equiv - {\cal Q}^2$ and $M_{\ell\bar\ell}^2 = (\ell_+ + \ell_-)^2$,
respectively. The Bjorken variable and the lepton energy loss is defined
conventionally as $\Bx \equiv {\cal Q}^2/(2 p_1 \cdot q_1)$ and $y \equiv
p \cdot q_1 / p \cdot k$. $\Delta^2 = (p_2 - p_1)^2$ is the $t$-channel
momentum transfer. The solid angle of the lepton pair in the photon rest
frame is $d {\mit\Omega}_\ell = \sin \theta_\ell d \theta_\ell d \varphi_\ell$.
$\phi$ is the azimuthal angle between the lepton and hadron scattering
planes and plays a distinguished role below. The amplitude ${\cal A}$
represents the sum of the signal in question and ``contaminating" Bethe-Heitler
processes,
\begin{equation}
{\cal A} = {\cal A}_{\rm VCS} + {\cal A}_{{\rm BH}_1} + {\cal A}_{{\rm BH}_2}
\, ,
\end{equation}
corresponding to the first, second and third diagrams in Fig.\ \ref{BHandDDVCS},
respectively. The latter two are expressed in terms of the Dirac and Pauli form
factors parametrizing the nucleon matrix element of the quark electromagnetic
current,
\begin{equation}
J_\mu =
\bar u_2
\left(
\gamma_\mu F_1 + \frac{i \sigma_{\mu\nu} \Delta_\nu}{2 M_N} F_2
\right) (\Delta^2)
u_1
\, ,
\end{equation}
with $u_i$ being the nucleon bispinor $u_i \equiv u (p_i)$. As we already
mentioned above, they are measurable elsewhere, see, e.g., \cite{Gay01} for
the most recent results.

The gauge invariant decomposition of the hadronic tensor contributing to
${\cal A}_{\rm VCS}$ was found in Ref.\ \cite{BelMul00} by an explicit
twist-three analysis at leading order of perturbation theory. To leading
twist accuracy, it reads
\begin{widetext}
\begin{eqnarray}
\label{ComptonTensor}
T_{\mu\nu}
\!\!\!&=&\!\!\! i \int d^4 z \, {\rm e}^{i q \cdot z}
\langle p_2 |
T \left\{ j_\mu (z/2) j_\nu (- z/2) \right\}
| p_1 \rangle
=
\frac{i}{2 p \cdot q} \epsilon_{\theta \lambda \rho \sigma} p_\rho \, q_\sigma
\left(
g_{\mu \theta} - \frac{p_\mu \, q_{2 \theta}}{p \cdot q_2}
\right)
\left(
g_{\nu \lambda} - \frac{p_\nu \, q_{1 \lambda}}{p \cdot q_1}
\right)
{\cal A}_1 (\xi, \eta, \Delta^2)
\nonumber\\
&-&\!\!\! \frac{1}{2} \left(
g_{\mu\nu} - \frac{q_{1 \mu} \, q_{2 \nu}}{q_1 \cdot q_2}
\right)
{\cal V}_1 (\xi, \eta, \Delta^2)
+
\frac{1}{2 p \cdot q}
\left(
p_\mu - \frac{p \cdot q_2}{q_1 \cdot q_2} \, q_{1 \mu}
\right)
\left(
p_\nu - \frac{p \cdot q_1}{q_1 \cdot q_2} \, q_{2 \nu}
\right)
{\cal V}_2 (\xi, \eta, \Delta^2)
\, .
\end{eqnarray}
\end{widetext}
Here the symmetrized combinations of momenta $q \equiv (q_1 + q_2)/2$ and
$p \equiv p_1 + p_2$ are used to define the generalized Bjorken variable
$\xi$ and skewedness parameter $\eta$:
\begin{eqnarray}
\label{Xi}
\xi \!\!&\equiv&\!\! -\frac{q^2}{p \cdot q} =\Bx
\frac{
{\cal Q}^2 - M_{\ell \bar\ell}^2 + \Delta^2/2
}{
\left( 2 - \Bx \right) {\cal Q}^2 - \Bx \left( M_{\ell \bar\ell}^2 - \Delta^2 \right)
}
\, , \\
\label{Eta}
\eta \!\!&\equiv &\!\! \frac{\Delta \cdot q}{p \cdot q} = - \xi \frac{
{\cal Q}^2 + M_{\ell \bar\ell}^2
}{
{\cal Q}^2 - M_{\ell \bar\ell}^2 + \Delta^2/2
} \, .
\end{eqnarray}

Since the light-cone dominance in Eq.\ (\ref{ComptonTensor}) is set by the average
virtuality $q^2$ at small $\Delta^2$ and moderate $\Bx$, the perturbative QCD
approach is applicable provided
\begin{equation}
\label{ApplicabilityPQCD}
|q^2|
\equiv
\left| {\cal Q}^2 - M_{\ell \bar\ell}^2 + \Delta^2/2 \right|/2
\gg 1 \, {\rm GeV}^2 \, .
\end{equation}
It is lost for ${\cal Q}^2 \sim M_{\ell\bar\ell}^2$, which implies, according to
Eq.\ (\ref{Xi}), that $\xi$ approaches zero. The minimal allowed value is set
by $|\xi_{\rm min}| \sim 1 \ {\rm GeV}^2/y s$, where $\sqrt{s}$ is the
center-of-mass energy. In the light-cone dominated region (\ref{ApplicabilityPQCD}),
the Compton form factors factorize into calculable coefficient functions and GPDs.
To leading order in coupling constant the Compton form factors satisfy the
generalized Callan-Gross relation
\begin{equation}
{\cal V}_2 = \xi {\cal V}_1 \, .
\end{equation}
Performing the Dirac decomposition one gets
\begin{eqnarray*}
{\cal V}_1 \!\!\!&=&\!\!
\bar u_2
\int \! d x \, C_- (x, \xi)
\left(
\gamma_+ H
+
\frac{i \sigma_{+\nu} \Delta_\nu}{2 M_N} E
\right) (x, \eta, \Delta^2)
u_1
, \\
{\cal A}_1 \!\!\!&=&\!\!
\bar u_2
\int \! d x \, C_+ (x, \xi)
\left(
\gamma_+ \gamma_5 \widetilde H
+
\frac{\gamma_5 \Delta_+}{2 M_N} \widetilde E
\right) (x, \eta, \Delta^2)
u_1
,
\end{eqnarray*}
in terms of GPDs $H$, $E$, $\widetilde H$ and $\widetilde E$ \cite{MueRobGeyDitHor94}.
The plus subscript stands for the contraction of the corresponding Lorentz index
with the light-like vector $n_\mu = - \xi (2 q_\mu + \xi p_\mu)/q^2$, which projects
out the leading power contribution. From here, one immediately sees the difficulty to
measure these functions: one of the dynamical variables enters integrated out with
the coefficient function, which reads to leading order in QCD coupling constant,
\begin{equation}
C_\mp (x, \xi) = \frac{1}{\xi - x - i 0} \mp \frac{1}{\xi + x - i 0} \, .
\end{equation}
One can get rid of the convolution provided the observable is sensitive to the
imaginary part of the Compton form factors only, i.e., $\Im{\rm m} {\cal V}_i$,
$\Im{\rm m}{\cal A}_1$.

The most illuminating experimental observables in this respect are single
beam or target spin asymmetries. Their advantage is that they (i) depend
linearly on Compton form factors, and (ii) are proportional to their imaginary
part. The complete result for the cross section (\ref{Xsection}), which is
represented by a double Fourier sum in azimuthal angles $d \sigma \sim
\sum_{m,n} \cos(m \phi) \{ cc_{m,n} \cos(m \varphi_\ell) + cs_{m,n}
\sin(m \varphi_\ell) \} + (\cos \leftrightarrow \sin)$, will be published
elsewhere. Here we limit ourselves to the case of the lepton helicity
difference and when one does not distinguish the final state pair with
respect to their angular distribution. The integration over the solid
angle ${\mit \Omega}_\ell$ leads to the vanishing of the interference term
${\cal A}_{\rm VCS} {\cal A}_{{\rm BH}_2}^\dagger$ and so the only term
which survives in the asymmetry $\Delta \sigma \equiv \sigma (\lambda = 1)
- \sigma (\lambda = - 1)$ is ${\cal A}_{\rm VCS} {\cal A}_{{\rm BH}_1}^\dagger$:
\begin{widetext}
\begin{eqnarray}
\label{SSAsection}
\frac{d \Delta \sigma}
{d \Bx d y d |\Delta^2| d \phi d M_{\ell \bar\ell}^2}
\!\!\!&=&\!\!\!
\frac{2 \alpha_{\rm em}^4}{3 \pi}
\frac{
(2 - y) y\sqrt{(1 - \Bx) {\cal Q}^2 - \Bx M_{\ell \bar\ell}^2} \sqrt{\Delta_{\rm min}^2 -
\Delta^2 } \sin \phi
}{
{\cal Q}^2 M_{\ell \bar\ell}^2 \Delta^2
\left(
\sqrt{1 - y} ( {\cal Q}^2 + M_{\ell \bar\ell}^2 )
+
2 (2 - y) \sqrt{(1 - \Bx) {\cal Q}^2 - \Bx M_{\ell \bar\ell}^2 }
\sqrt{\Delta^2_{\rm min} - \Delta^2} \cos\phi
\right)
}
\nonumber\\
&\times&\!\!\!\left(
F_1 (\Delta^2) H_s (\xi, \eta, \Delta^2)
+
\xi (F_1 + F_2) (\Delta^2) \widetilde H_s (\xi, \eta, \Delta^2)
-
\frac{\Delta^2}{4 M_N^2} F_2 (\Delta^2) E_s (\xi, \eta, \Delta^2)
\right)
\, ,
\end{eqnarray}
\end{widetext}
where $F_s (\xi, \eta, \Delta^2) \equiv F (\xi, \eta, \Delta^2) -
F (- \xi, \eta, \Delta^2)$. The minimal value of the $t$-channel momentum
transfer is $\Delta^2_{\rm min} \approx - 4 M_N^2 \eta^2/(1 - \eta^2)$.
The unique feature of the process is that GPDs can be studied as functions
of all three arguments independently. Note that if the timelike virtuality
is away from the resonance region, the hadronic component of the photon can
be neglected and consequently all subprocesses with vector meson production.

\begin{figure}[htb]
\begin{center}
\mbox{
\begin{picture}(0,165)(90,0)
\put(0,0){\insertfig{6}{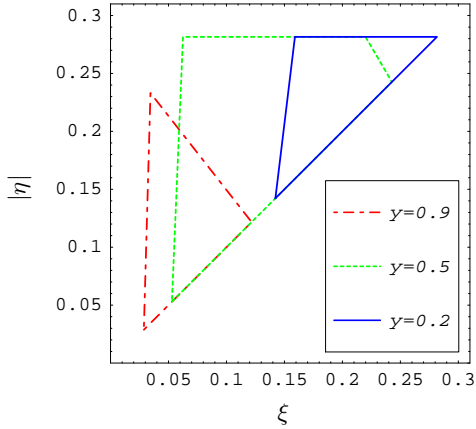}}
\put(90,1){$\xi$}
\put(-12,84){\rotate{$|\eta|$}}
\end{picture}
}
\end{center}
\vspace*{-0.3cm}
\caption{\label{XiEtaPlane} The $\xi$-$\eta$ coverage with $E = 11 \, {\rm GeV}$
electron beam. Only the regime ${\cal Q}^2 > M_{\ell\bar\ell}^2$ is shown. The
kinematics is taken as follows: $\Delta^2 = - 0.3 \, {\rm GeV}^2$, ${\cal Q}^2$
varies from 1 GeV$^2$ to 4 GeV$^2$ and $M_{\ell\bar\ell}^2$ from 0 to ${\cal Q}^2
- 1 \, {\rm GeV}^2$. The perimeters of areas covered for different lepton energy
losses $y$ are shown as described in the legend.}
\end{figure}

As Fig.\ \ref{XiEtaPlane} demonstrates, the region of the GPD surface accessible
in this reaction for a fixed target experiment setup is quite extensive. For a
given $\xi$, cf. Eq.\ (\ref{Xi}), with $\Bx = {\cal Q}^2/(2 M_N E y)$, the
variable $\eta$ varies in limits
\begin{equation}
{\rm min} \, |\eta|
\leq | \eta | \leq
{\rm min}
\left\{
{\rm max} \, |\eta|, |\eta_{\rm cut}|
\right\} \, ,
\end{equation}
where $\eta$ is evaluated according to Eq.\ (\ref{Eta}) and the bound
$|\eta_{\rm cut}| \equiv \sqrt{- \Delta^2/(4 M_N^2 - \Delta^2)}$ comes from
the condition $|\Delta^2| \geq |\Delta_{\rm min}^2|$. The minimal value of
$\xi$ is governed by the requirement of applicability of perturbative QCD
treatment, Eq.\ (\ref{ApplicabilityPQCD}), chosen here as ${\cal Q}^2 \geq
M_{\ell\bar\ell}^2 + 1 \, {\rm GeV}^2$. The restriction $|\eta| \geq \xi $
is a simple consequence of the timelike nature of the final state photon.

The ratio of (\ref{SSAsection}) to the DVCS signal
\begin{eqnarray}
\frac{1}{d \Delta \sigma_{\rm DVCS}}
\int_{4 m_e^2}^{\sim {\cal Q}^2} d M_{\ell \bar\ell}^2 \,
\frac{d \Delta \sigma}{d M_{\ell \bar\ell}^2}
\sim
\frac{\alpha_{\rm em}}{3 \pi}
\ln \frac{{\cal Q}^2}{m_e^2}
\, ,
\end{eqnarray}
is of order $0.01$ for ${\cal Q}^2 \simeq 2 \, {\rm GeV}^2$. Difficulties
in measuring such a small cross section will be overcome at high-luminocity
machines, like JLab@12GeV with $L = 10^{35}{\rm cm}^{-2} {\rm s}^{-1}$ or
in collider experiments due to growth of the cross section with increasing $\xi$.

\begin{figure}[htb]
\begin{center}
\mbox{
\begin{picture}(0,245)(90,0)
\put(0,120){\insertfig{7}{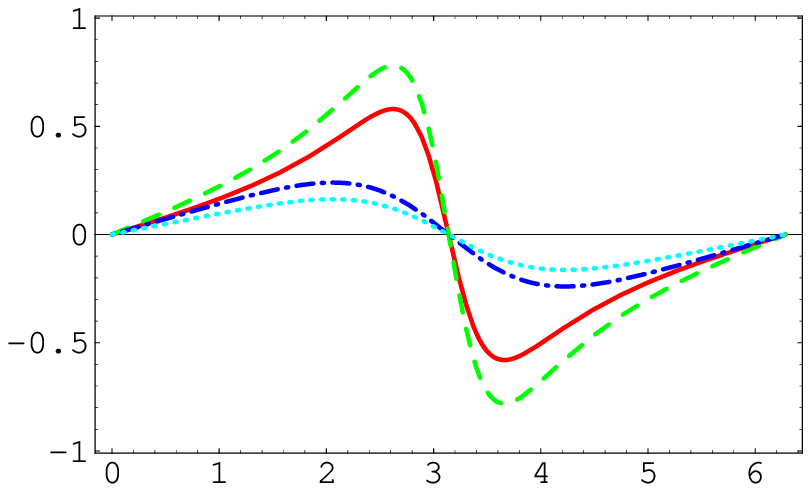}}
\put(0,0){\insertfig{7}{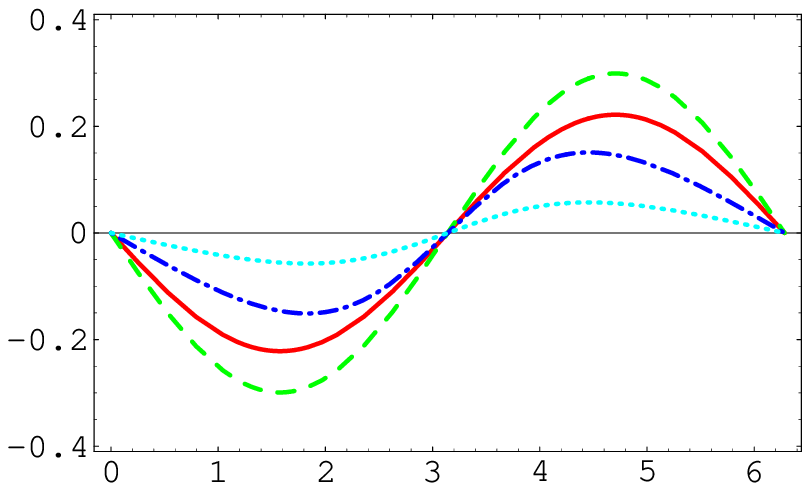}}
\put(115,213){\fbox{$\!\!\begin{array}{l}
\Bx = 0.2 \\ {\cal Q}^2 = 2 \ {\rm GeV}^2 \\ \Delta^2 = - 0.15 \ {\rm GeV}^2
\!\!\end{array}$}}
\put(-25,125){\rotate{
$M^2_{\ell \bar\ell}
d \Delta \sigma
/
d \Bx d y d |\Delta|^2 d \phi d M^2_{\ell \bar\ell}
$}}
\put(-13,163){\rotate{$[{\rm pb}/{\rm GeV}^2]$}}
\put(115,32){\fbox{$\!\!\begin{array}{l}
y = 0.5 \\ M^2_{\ell \bar\ell} = 2 \ {\rm GeV}^2 \\ \Delta^2 = - 0.15 \ {\rm GeV}^2
\!\!\end{array}$}}
\put(-25,-2){\rotate{
$\sqrt{{\cal Q}^2}
d \Delta \sigma
/
d \Bx d y d |\Delta|^2 d \phi d M^2_{\ell \bar\ell}
$}}
\put(-13,45){\rotate{$[{\rm pb}/{\rm GeV}^3]$}}
\put(100,-10){$\phi [{\rm rad}]$}
\end{picture}
}
\end{center}
\caption{\label{Fig-JLAB-1}
Azimuthal angle dependence of the scaled cross section of $e^- p \to e^- p
e^- e^+$ for spacelike (top) and timelike (bottom) $q^2$ at $E = 11 \
{\rm GeV}$. The solid, dash-dotted and dashed, dotted curves represent
predictions for FPD and DD (with $b = 1$, $B_{\rm sea} = 9 \ \mbox{GeV}^{-2}$)
models of Ref.\ \protect\cite{BelMulKir01}, respectively. Top:
$M^2_{\ell \bar\ell} = 0$ for solid, dashed and $M^2_{\ell \bar\ell}
= 0.7 \ \mbox{GeV}^2$ for dashed, dash-dotted curves, respectively.
Bottom: ${\cal Q}^2 = 0$ for solid, dashed and ${\cal Q}^2 = 1\ \mbox{GeV}^2$
for dash-dotted, dotted curves, respectively.
}
\end{figure}

\begin{figure}[t]
\vspace{-0.2cm}
\begin{center}
\mbox{
\begin{picture}(0,120)(120,0)
\put(0,0){\insertfig{4}{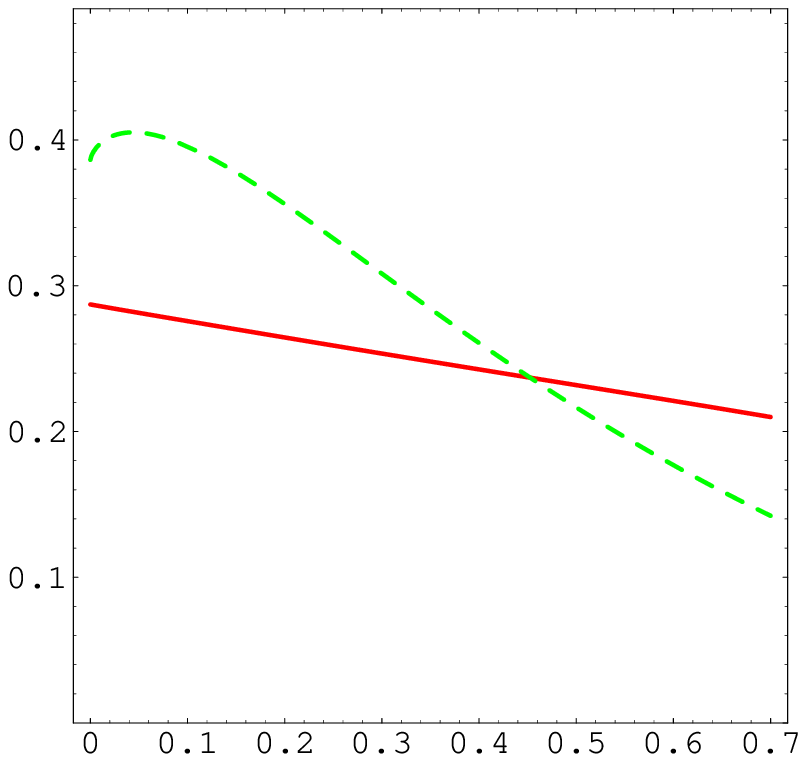}}
\put(125,0){\insertfig{4}{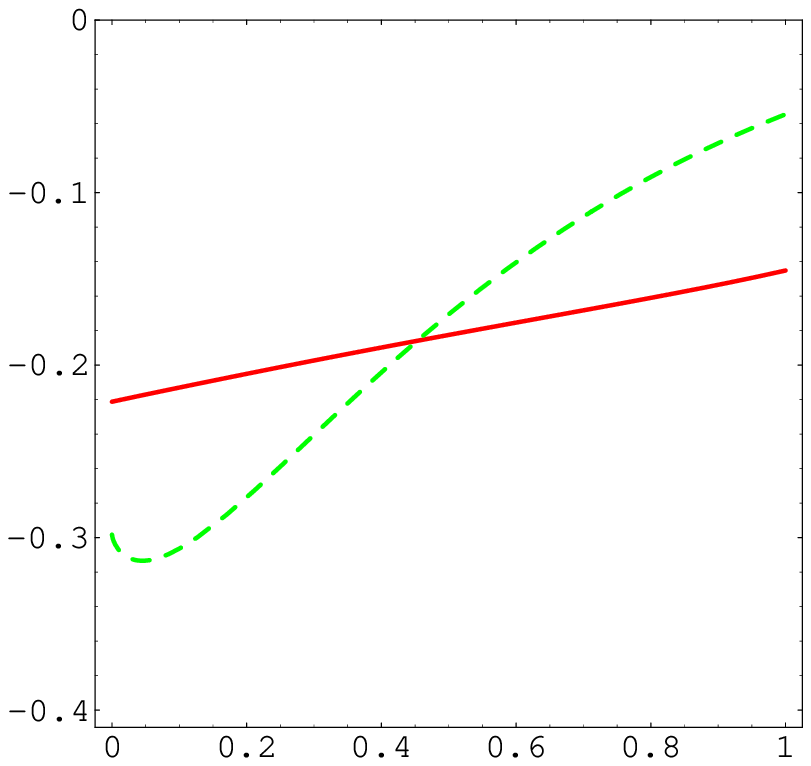}}
\put(40,-10){$M_{\ell\bar\ell}^2 [{\rm GeV}^2]$}
\put(170,-10){${\cal Q}^2 [{\rm GeV}^2]$}
\end{picture}
}
\end{center}
\caption{\label{Fig-JLAB-2}
The dependence of scaled cross section for FPD (solid) and DD (dashed)
models at $\phi = \pi/2$ for $M_{\ell\bar\ell}^2 d \Delta \sigma$ (left)
and $\sqrt{{\cal Q}^2} d \Delta \sigma$ (right) for the same kinematical
settings as in Fig.\ \ref{Fig-JLAB-1}, top and bottom panels, respectively.
}
\end{figure}

\begin{figure}[t]
\vspace{-0.2cm}
\begin{center}
\mbox{
\begin{picture}(0,115)(120,0)
\put(0,0){\insertfig{4}{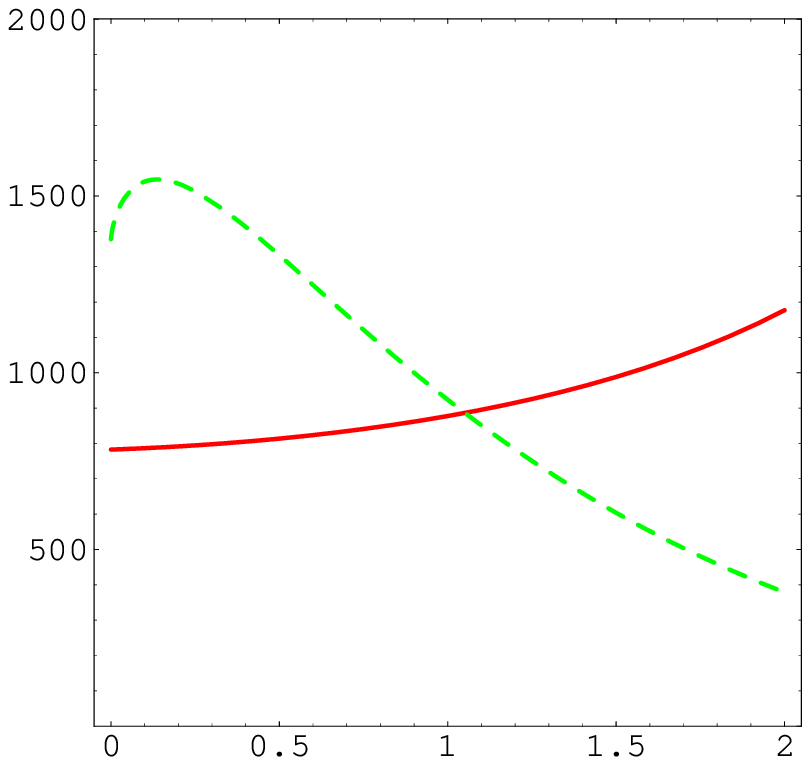}}
\put(125,0){\insertfig{4}{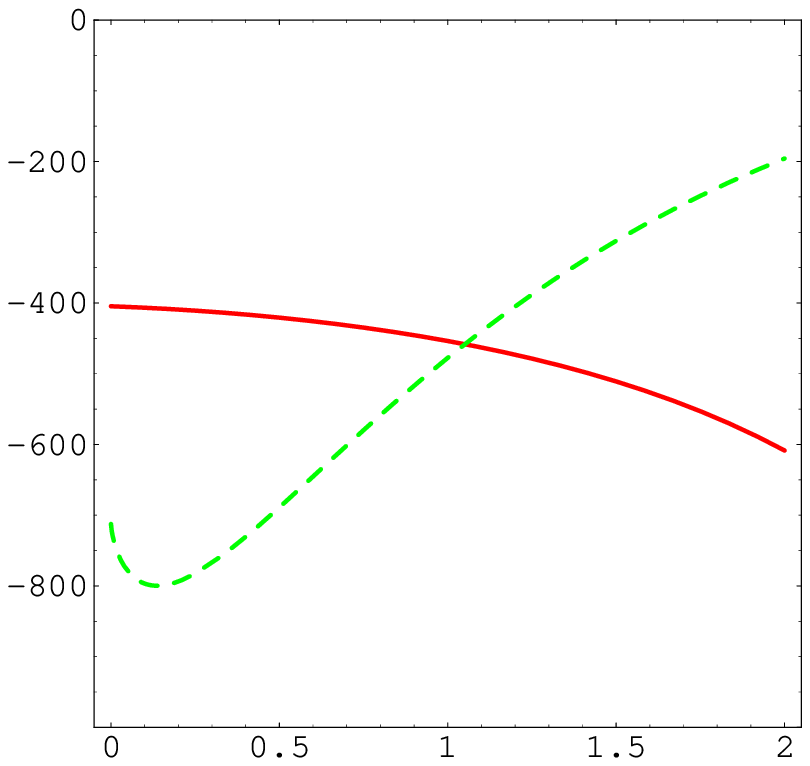}}
\put(40,-10){$M_{\ell\bar\ell}^2 [{\rm GeV}^2]$}
\put(170,-10){${\cal Q}^2 [{\rm GeV}^2]$}
\end{picture}
}
\end{center}
\caption{\label{Fig-HERA}
Same as in Fig.\ \ref{Fig-JLAB-2}, displayed for HERA kinematics with
$y = 0.45$, $\Delta^2 = - 0.1 \ \mbox{GeV}^2$, and ${\cal Q}^2 = 4 \
\mbox{GeV}^2$ (left) and $M^2_{\ell \bar\ell} = 4 \ \mbox{GeV}^2$ (right).}
\end{figure}

The cross section (\ref{SSAsection}) being directly proportional to GPDs is
extremely sensitive to their skewedness dependence. Such kinds of measurements
can easily establish the credibility of our current understanding of
mesonic-like components of GPDs and thus constrain the region which is
lacking in the establishment the total spin sum rule \cite{Ji96}. In Fig.\
\ref{Fig-JLAB-1}, we present estimates for JLab@12GeV kinematics using two
different models of GPDs: A model without skewedness dependence, which
corresponds to the conventional parton density taken to be the same at all
values of $\eta$, the so-called forward parton distribution (FPD) model.
And a model based on a more sophisticated construction \cite{Rad99} which
takes the aforementioned parton densities as input and leads to a nontrivial
$\eta$-dependence. It is called the double distribution (DD) model. For
specific details we refer to Ref.\ \cite{BelMulKir01}. The most prominent
way to discriminate between these models is to study the $M_{\ell\bar\ell}^2$
or ${\cal Q}^2$ dependence of the cross section (\ref{SSAsection}) as
shown in Fig.\ \ref{Fig-JLAB-2}. One will be easily able to distinguished
between different behaviors by varying the timelike photon virtuality over a
short interval below the $\rho$-meson threshold. As Fig.\ \ref{Fig-HERA}
shows, the higher energy of HERA results into even higher sensitivity to
the skewedness dependence of GPDs. We like to mention that at HERA the
skewedness dependence can already be studied in measurements of the
unpolarized cross section, since the imaginary part of the Compton
amplitude dominates over the real part.

To conclude, the exclusive electroproduction of lepton pairs provides a
unique opportunity to determine exhaustive information on the nucleon's
internal structure by accessing inter-parton correlations. The underlying
GPD can be mapped uniquely as a function of all its variables which encode
dynamics in longitudinal and transverse spaces. Our analysis demonstrates
a high sensitivity of the lepton helicity asymmetry to the skewedness
dependence of GPDs. Similar conclusions apply to nucleon spin asymmetries
which again extract the interference term of Bether-Heitler and Compton
amplitude and separate yet another combination of functions in question.
The angular distribution of the final state lepton pairs has a very rich
structure and its measurement will lead to an indispensable complementary
information on GPDs. One might expect that such studies at collider energies
will pin down the skewedness dependence of gluon GPDs in the small-$\xi$
region and thus reduce theoretical uncertainties in diffractive leptoproduction
of vector mesons. Finally, let us point out that the difference between the
space- and timelike region is perturbatively computable and so the onset of
the light-cone dominance can be elucidated as well.

We would like to thank L.~Elouadrhiri, S.~Stepanyan and M.~Diehl for
discussions. Both authors are grateful to the Theory Group and Hall B
at Jefferson Lab and D.M. thanks the Nuclear Theory Group at the
University of Maryland for the hospitality at the intermediate stages
of the work. This work was supported by the US Department of Energy
under contract DE-FG02-93ER40762.


\end{document}